\documentclass[fleqn,10pt,onecolumn,twoside]{wlscirep}

\usepackage{subcaption}
\usepackage{tabularx}
\usepackage{xcolor}

\title{Laser induced strong-field ionization gas jet tomography}

\author[1]{Oshrat Tchulov}
\author[2]{Matteo Negro}
\author[2]{Salvatore Stagira}
\author[2]{Michele Devetta}
\author[2]{Caterina Vozzi}
\author[1,*]{Eugene Frumker}

\affil[1]{Department  of Physics, Ben-Gurion University of the Negev, Beer-Sheva 84105, Israel}
\affil[2]{IFN-CNR and Dipartimento di Fisica-Politecnico di Milano, Piazza Leonardo da Vinci 32, 20133 Milano, Italy}
\affil[*]{Correspondence: E. Frumker, Email: efrumker@bgu.ac.il}


\begin{abstract}
We introduce a novel in-situ strong field ionization tomography approach for characterizing the spatial density distribution of gas jets.
We show that for typical intensities in high harmonic generation experiments,
the strong field ionization mechanism used in our approach provides an improvement in the resolution close to factor of 2 (resolving about 8 times smaller voxel volume), when compared to linear/single-photon imaging
modalities.
 We find, that while the depth of scan in linear tomography is limited by resolution loss due to the divergence of the driving laser beam, in the proposed approach  the depth of focus is localized due to the inherent physical nature of strong-field interaction and discuss implications of these findings. We explore key aspects of the proposed method and compare it with commonly used single- and multi-photon imaging mechanisms. The proposed method will be particularly useful for strong field and attosecond science experiments.
\end{abstract}
\begin{document}

\flushbottom
\maketitle

\thispagestyle{empty}

\section*{Introduction}

A key ingredient in attosecond technology is the generation of attosecond optical pulse trains \cite{Paul_Rabbit_Science2001} and isolated attosecond pulses \cite{Sansone_isolated_atto_Science_2006}. Currently, the dominant technique for attosecond pulse generation is based on strong field interaction of a driving laser field with gas targets. Typically, an intense femtosecond laser pulse is focused on an atomic or a molecular gas jet. The highly nonlinear  electronic response of the medium to the periodic strong driving field, causes high harmonics generation (HHG) of the laser field \cite{corkum1993plasma}. In the case of ideal phase-matching, emission from individual atoms or molecules is coherently added in phase, and the total harmonic intensity is therefore proportional to the square of the emitters' number. However, in reality, there is a delicate interplay between the phases of generated and driving fields \cite{Constant_optimizing_PRL1999} that influence the coherent build-up of HHG radiation.  Experimentally, this implies, that the efficiency, temporal and spatial profile of the HHG and attosecond pulse generation, depends on both the driving laser spatial and temporal field distribution, as well as on the spatial density distribution of the generating medium itself.
Therefore, detailed experimental characterization of the generating medium density distribution is essential for optimization of the attosecond pulse generation process, and the acquisition of a more profound understanding of the underlying physical phenomena \cite{frumker2012order}.%

 Tomography is an effective non-destructive technique, which provides 3D image of the internal structure of materials. It is being widely used in medicine \cite{huang1991optical}, seismology \cite{shapiro2005high}, materials science \cite{salvo2003x,baruchel2000x} and many other  scientific and industrial areas. Tomographic reconstruction techniques based on linear  or perturbative non-linear interaction also have been used to map the spatial density distribution of gas jets \cite{malka2000charact, Failor_fluorescence_gas_tomogr_RSI_2003, thorpe2009tomography, schofield2009absolute, wachulakextreme}.
 %
 %
 %
 %
 %
 In this paper we present a novel in-situ method, based on strong-field ionization mechanism, to measure the spatial distribution of a generating medium density itself.

\section*{Methods}

\begin{figure}[hbpt]
    \begin{center}
	\fbox{\includegraphics[width=\linewidth]{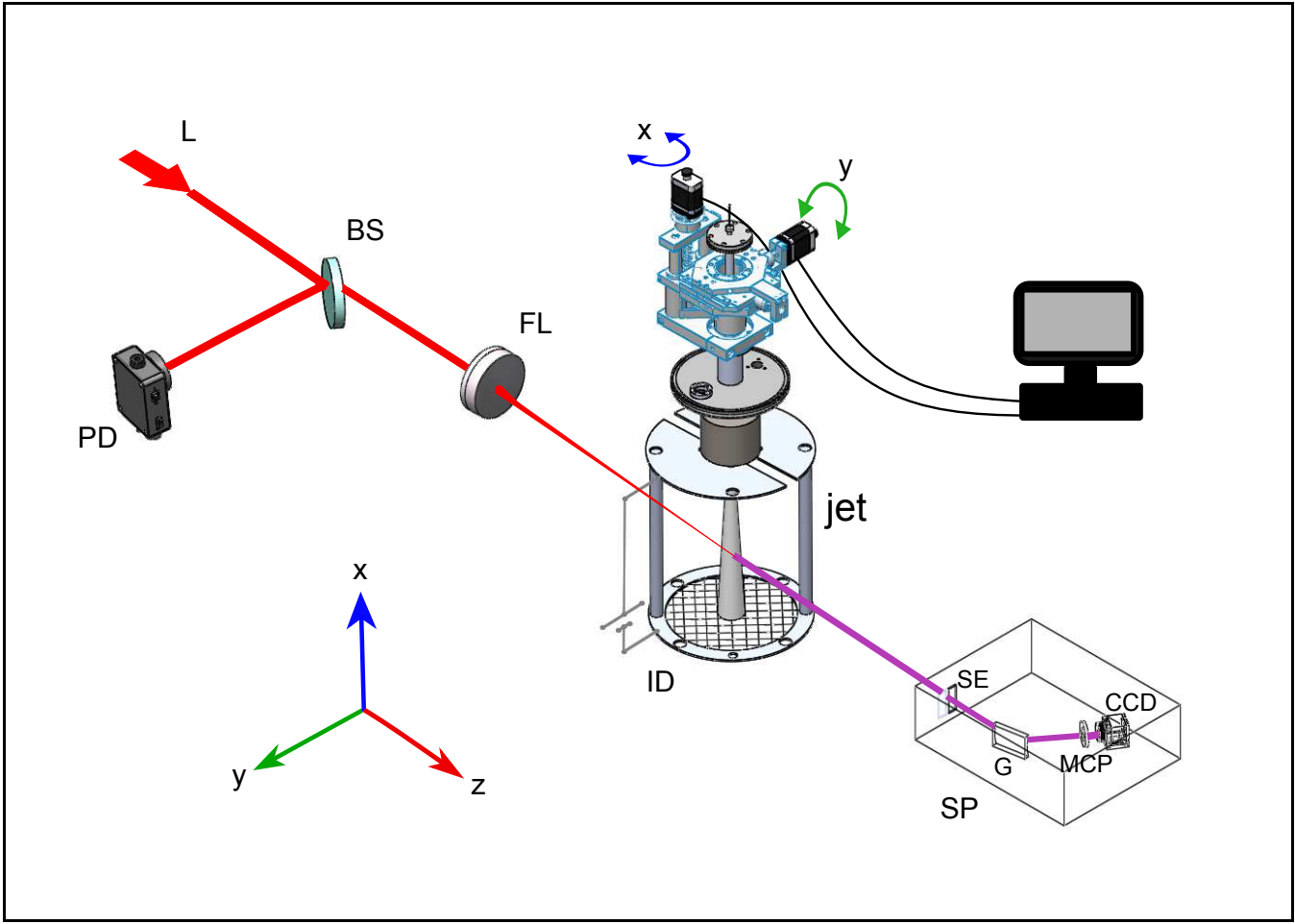}}

    \captionsetup{justification=justified}

    \caption{Schematic of the experimental setup. L-laser; BS-beam splitter; PD-photodiode detector; FL-focusing lens; ID-ion detector; SP-spectrometer; SE-spectrometer slit; G-grating.}
    \label{fig:system}
    \end{center}
\end{figure}

  The schematic of the proposed approach is illustrated in Fig. \ref{fig:system}. %
An intense femtosecond laser beam is focused onto a gas jet, which is the HHG source.  A spectrometer (SP)
is used to measure the spectrally resolved high harmonics signals generated by the gas jet, and an ion detector (ID) \cite{shiner_wavelength_scaling_2009} to record the gas ions' signal, which accompany the HHG process. The laser beam is scanned along the x- and the y-axes in order to build tomographic signal data for the reconstruction of the gas jet density map.
 In our proof-of-principle experiment we used a gas jet with cylindrical symmetry, so that the x-y scan is sufficient to collect a complete data set for the tomographic reconstruction. In the general case of an asymmetric jet, one will have to collect the data by rotational scanning of the jet relative to the driving laser beam.

 For each scanned position in the x-y plane, three measurements are taken: the spectrum of the HHG signal, the ion signal, and the laser intensity signal. The tomographic reconstruction  can be obtained separately from either HHG signal or from the ion signal. In this article we will focus on the ionization signal.


  %
%
%
%
%

 HHG can be modeled as a three-step semiclassical process \cite{corkum1993plasma,schafer1993above}, in which within one optical cycle of an intense laser field (i) an electron is ionized from an atom or a molecule, (ii) the released electron accelerates in the continuum and (iii) finally the electron recombines with its parent ion.
However, only a small fraction of ionized electrons in the first step makes it through all three steps until recombination. Thus, after the driving laser pulse is gone, an ionized gas medium is left behind. The ionization signal measured by the ion detector is proportional to the density of the gas within the laser focal volume. In this work we exploit this ion signal, driven by the strong laser field ionization mechanism, for our measurements.

The first building block of the three-step model is nonlinear ionization of the atomic system by an optical field, whose frequency, $\omega_L$, is lower than the ionization energy $I_p$ ($\omega_L\ll I_p$ ) of the outermost electron in the system (keeping the tradition, we use atomic units).

The fundamental insight into this strong field ionization process was provided by Keldysh \cite{keldysh1965ionization}.
  According to Keldysh's approach, the physics of the ionization process is determined by the adiabaticity (Keldysh) parameter, $\gamma=\omega_L\sqrt{2I_p}/E_L$, where $E_L$ is the laser's electric field. Hence, the two limiting regimes of the universal process of nonlinear ionization are identified. In the multiphoton ($\gamma\gg 1$) regime, the dependence of the ionization rate, $W_i$, on the electric field of the optical wave is a power law: $W_i\propto E_L^{2K}$, where $K=\lfloor I_p/\omega_L+1 \rfloor$ is the threshold number of absorbed photons required by energy conservation. In the tunnel-ionization  ($\gamma\ll 1$) regime, the nonlinear ionization rate depends exponentially on $E_L$: $W_i\propto exp(-2E_a/(3E_L))$, where $E_a=(2I_p)^{3/2}$ is the characteristic atomic electric field strength. The rate formula in this case can be obtained by averaging the equation for tunneling ionization in a constant electric field over half cycle of the alternating electric field of the optical wave\cite{delone2012multiphoton}.

Shortly after the appearance of Keldysh's work its results were refined in the Perelomov-Popov-Terentev  (PPT) model \cite{perelomov1966ionization} for a short range potential and arbitrary values of the Keldysh parameter $\gamma$. The effect of the Coulomb interaction between the ejected electron and the atomic core was taken into account in a subsequent paper \cite{perelemov1967ionization} through a first order correction in the quasi-classical action. This correction is applicable for not too weak optical fields or too large values of the Keldysh parameter \cite{perelemov1967ionization,perelomov1968allowance}. It was shown experimentally, for most of the noble elements, that the accuracy of the PPT model is excellent up to $\gamma=3-4$. \cite{larochelle1998coulomb,yudin2001nonadiabatic}. Another commonly used model for calculating the ionization rate is the so-called Ammosov-Delone-Krainov  (ADK) model \cite{ammosov1986tunnel}. This model is based on the tunneling ionization rate equation given in \cite{perelomov1966ionization}, which is strictly valid only in the limit $\gamma\ll 1$. It was, however,  shown numerically for several atoms, that this model coincides with the more general PPT model for values of $\gamma\lesssim 0.5$. As a result, the ADK model is frequently used as well for intermediate values of $\gamma$, up to $\gamma\sim 0.5$ \cite{yudin2001nonadiabatic}.%
%
%
%
%
%

\section*{Results and Discussion}

 First, we consider both the ADK and PPT models to explore major characteristics of the laser induced strong-field ionization tomography, and compare it to the commonly used fluorescence tomography \cite{Failor_fluorescence_gas_tomogr_RSI_2003} (where single excitation photon is converted into one emitted photon at lower energy), and to the widely studied perturbative nonlinear imaging modalities \cite{Hellwarth_SHG_microscopy_AO_1975, Gannaway_SHG_microscopy_OQE_1978, Barad_THG_microscopy_APL_1997, schofield2009absolute}.

\begin{figure}[hbpt]
    \begin{center}
	\includegraphics[width=\linewidth]{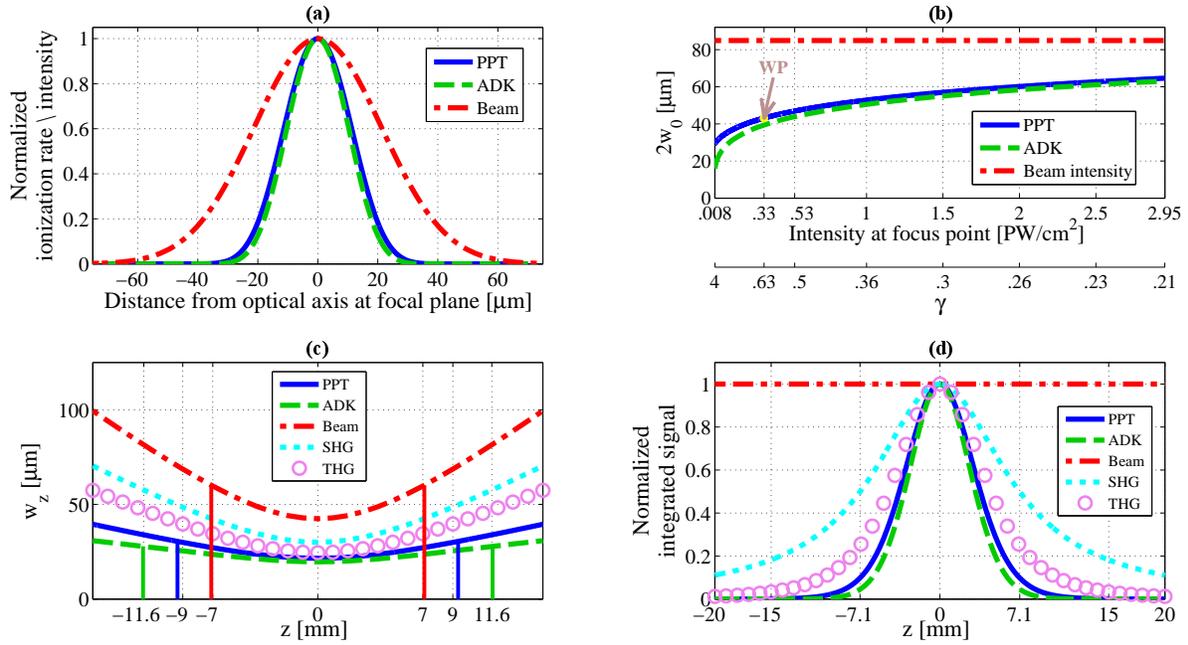}

    \captionsetup{justification=justified}

\caption{Strong field ionization (PPT \& ADK models) driven by a focused Gaussian beam:
(a) PPT (solid blue) and ADK (dashed green) ionization rate profiles generated by a focused  Gaussian beam with the intensity profile shown as red dash-dotted curve;
(b)The waist diameter as a function of the laser intensity at the focus point (Our experimental working point is indicated by WP);
(c) Ionization rate width $\textrm{w}_z$ (defined at the  $1/e{^2}$ of maximum ionization)  of PPT and ADK signals along the optic axis, shown together with SHG,THG and the laser beam width. Solid vertical lines show the distance from the focal plane where $\textrm{w}_z$ increases by factor $\sqrt{2}$ in analogy with the Rayleigh range ($z_R$);
(d) Signal integral across the plane perpendicular to the optical axis under the assumption of constant gas density. Note the variation of the integrated ion signal vs. constant integral of the laser intensity (i.e. power) along the propagation direction.
}
\label{fig:ratesAndBeam}
\end{center}

\end{figure} %

Figure \ref{fig:ratesAndBeam}\textcolor{blue}{a} shows calculated ADK and PPT ionization rate profiles generated by a focused Gaussian beam.
For Fig. \ref{fig:ratesAndBeam}, peak intensity  of $\sim3.3\times10^{14} \textrm{W}/\textrm{cm}^2$ and beam width, $\textrm{w}_0=42 \mu m $, of the driving laser were assumed, which correspond to our proof-of-principle experiment.
The spatial resolution of the strong field ionization tomographic reconstruction is fundamentally limited by the width of this ionization profile, as the fluorescence tomography \cite{Failor_fluorescence_gas_tomogr_RSI_2003} would be limited by the width of the
laser beam intensity.
In analogy with the width definition of a Gaussian laser beam intensity, we define the width of the ionization profile as the distance from the optical axis (i.e. where maximum ionization rate occurs) to the point where ionization rate decreases by a factor of $e^{-2}$.
Our calculations show that the widths of the ionization profiles (for both ADK and PPT models)are narrower, by at least a factor of 2, than  the width of the driving laser beam. It means at least twice improvement in resolution or, using the language of tomography, the minimum resolvable voxel \cite{Amanatides_voxel_Eurographics_1987} volume is 8 times smaller as compared to the fluorescence tomography (or any other linear tomography technique).

\begin {table}[hbpt]
\caption{Comparison between the spatial dependence (in the focal plane) of common imaging modalities and strong field ionization approach. The extra factor, $g(\gamma)$, in the PPT exponential, is a function which gets the value 1 for $\gamma=0$, and decreases monotonically with increasing $\gamma$, causing the width of the PPT curve to be somewhat wider than the ADK curve.}
\centering
{
	\renewcommand{\arraystretch}{2.5}
	\begin{tabularx}{\linewidth} {c|c|c}
	\hline\hline
	Signal type & Transverse variation & Waist radius \tabularnewline
	\hline
	
	Fluorescence & $\propto I\sim e^{-2(\frac{r}{\textrm{w}_0})^2}$ & $\textrm{w}_0$\tabularnewline
	
	SHG & $\propto I^2\sim e^{-2(\frac{r}{\textrm{w}_0/\sqrt{2}})^2}$ & $\dfrac{\textrm{w}_0}{\sqrt{2}}\approx 0.7 \textrm{w}_0$\tabularnewline

		
	THG & $\propto I^3\sim e^{-2(\frac{r}{\textrm{w}_0/\sqrt{3}})^2}$ & $\dfrac{\textrm{w}_0}{\sqrt{3}}\approx 0.58 \textrm{w}_0$\tabularnewline
	

    PPT & $\propto exp\{-\frac{2}{3}E_a e^{(\frac{r}{\textrm{w}_0})^2}g(\gamma)\}$  & $0.5\textrm{w}_0$\tabularnewline

    ADK & $\propto exp\{-\frac{2}{3}E_a e^{(\frac{r}{\textrm{w}_0})^2}\}$  & $0.46\textrm{w}_0$\tabularnewline

	\hline
	\end{tabularx}
}
\label{tab:FluorescenceSpacialResolution}
\end{table} %

It is interesting to compare explicitly the expected resolution of different laser based imaging modalities, as shown in Table \ref{tab:FluorescenceSpacialResolution}.
 In fluorescence tomography \cite{Failor_fluorescence_gas_tomogr_RSI_2003}, a single excitation photon is converted into one
 emitted (fluorescence) photon at a lower energy. This is a linear process, implying, as mentioned, that the spatial resolution of the fluorescence tomography is limited by the width  ($\textrm{w}_\textrm{0}$) of the driving laser intensity.
 There is a tremendous research interest focused on a variety of nonlinear microscopy techniques, such as second harmonic generation (SHG) \cite{Hellwarth_SHG_microscopy_AO_1975,Gannaway_SHG_microscopy_OQE_1978}, two-photon fluorescence (TPF) \cite{Denk_2photon_microscopy_Science_1990}, third harmonic generation (THG)\cite{Barad_THG_microscopy_APL_1997}, Raman  \cite{Freudiger_stim_Raman_microscopy_Science_2008}, and other microscopies. The common feature of these techniques is that all of them are based on either second order or third order perturbative nonlinear processes. The major motivation for using these nonlinear microscopy techniques is resolution improvement.
  As shown in Table \ref{tab:FluorescenceSpacialResolution} while second order and third order perturbation imaging techniques provide resolution improvement resulting in an effective spot size of $\textrm{w}_\textrm{0}/\sqrt(2)\simeq 0.7 \textrm{w}_\textrm{0}$ and  $\textrm{w}_\textrm{0}/\sqrt(3)\simeq 0.58 \textrm{w}_\textrm{0}$ respectively, the strong field nonlinear interaction, that is at the heart of our approach, improves the resolution by at least a factor of 2 for typical conditions used in our experiment and calculations.


Now, we turn our attention to the resolution dependence on the  driving laser field intensity.
In a perturbative nonlinear process of order $n$ , the generated nonlinear signal
is proportional to $I^n\propto \exp[-2(\sqrt(n)r/\textrm{w}_\textrm{0})^2]$.
Thus, the expected resolution -  $\textrm{w}_\textrm{0}/\sqrt(n)$ is independent of the driving intensity, under the general restriction of a perturbative intensity regime.
On the other hand, in the case of strong field nonlinear ionization interaction, the situation is more subtle since there is an inherent physical mechanism that couples the peak intensity and the expected resolution, as follows from the analytical formulas in Table \ref{tab:FluorescenceSpacialResolution} corresponding to the ADK and PPT signals and plotted in Fig. \ref{fig:ratesAndBeam} (for a  driving laser width of $2\textrm{w}_\textrm{0}=85 \mu m$).
One can see that both ionization models produce very similar results, namely, the resolution improvement gradually decreases
as the driving intensity increases. For the range of intensities optimal for HHG (shown as WP on Fig.\ref{fig:ratesAndBeam}\textcolor{blue}{b}), we obtain an improvement in resolution by a factor of about 2.


Another important parameter to consider is the tomography depth range, which is possible to achieve along the laser beam propagation direction.
There are two factors that can limit this range - (1) the divergence of the generated signal along the propagation direction (analogous to the Rayleigh range in the case of linear tomography), which reduces the resolution, and (2) the reduction in integrated signal strength, generated across the plane perpendicular to the optical axis, with the distance from the focal point along the propagation direction.

Concerning the first limitation factor, Fig. \ref{fig:ratesAndBeam}\textcolor{blue}{c} compares the
strong field ionization rate width, $\textrm{w}_z$  of PPT and ADK signals, along the optic axis to the corresponding widths of the perturbative SHG and THG signals, as well as to the driving laser beam width.
One can see that the divergence of the ionization signals is not only smaller than
the divergence of the driving laser beam (assuming a fundamental Gaussian mode), but also smaller even compared to the divergence of SHG and THG signals. In other words, the equivalent of the Rayleigh range for ionization is larger than the Rayleigh ranges of the driving laser beam and of the perturbative signals.
For example, in our case, while the Rayleigh range of the driving laser is 7.1mm, the ion signal divergence length for the PPT and ADK models is 9.3mm and 11.6mm respectively, as shown in Fig. \ref{fig:ratesAndBeam}\textcolor{blue}{c}.

Regarding the second limitation factor, while the integrated signal generated across the plane perpendicular to the optical axis remains constant for the single photon tomography modality, it decreases with increasing distance from the focal plane, in the case of the perturbative signals, and even more so in the case of strong field non-linear interaction, as shown in Fig. \ref{fig:ratesAndBeam}\textcolor{blue}{d}. In a way, this is analogous to the effect that happens in the perturbative non-linear microscopy  and enables the inherent  localization of excitation (so called "optical sectioning") \cite{Denk_2photon_microscopy_Science_1990}. In the field of multiphoton microscopy, this localization of excitation is often quoted as the key advantage compared to single photon excitation microscopy techniques \cite{Denk_2photon_microscopy_Science_1990}.

However, in our context of strong field tomography this ionization localization phenomena may have several important implications. First, if the spatial extent of the measured gas jet is smaller than the localization width (as happened to be in our proof-of-principle experiment), we may just use our data without any special processing needs, as of usual tomographic reconstruction data set.
Second, if the spatial extent of the measured gas jet is comparable to the localization width, the  tomographic reconstruction data set has to be normalized according to a normalization curve.
This curve can be either calculated from the strong field ionization models, as shown in Fig. \ref{fig:ratesAndBeam}\textcolor{blue}{d}, or measured
experimentally, scanning the narrow (as compared to the localization width) gas jet along the driving laser beam.
It is important to note,  that the localization width (Fig. \ref{fig:ratesAndBeam}\textcolor{blue}{d}) is smaller than the divergence of the ionization signal (Fig. \ref{fig:ratesAndBeam}\textcolor{blue}{c}).
 Thus, we may conclude that in the strong field ionization tomography, the leading mechanism that poses the tomography
 depth limit is the strong field ionization localization. This is in contrast to single-photon modalities, where the tomography depth limit is given by the divergence (confocal parameter) of the driving laser field.

Third, this inherent strong field ionization localization can be an advantage, if the gas jet is injected into a system with some background/buffer gas environment. In such a case, single photon tomography \cite{Failor_fluorescence_gas_tomogr_RSI_2003} methods would produce strong background signal that would increase noise and reduce the dynamic range of the measurement. The inherent localization phenomena in the strong field ionization tomography will resolve this problem, as the background gas will not be ionized and the tomography measurement will be free of this background and noise contribution.

In our proof-of-principle experiment, a 60 fs Ti-Sapphire amplifier output was passed through a hollow fiber and a gas filamentation cell to improve beam quality and to broaden its spectrum. The output of the filamentation was further compressed down to 15 fs by a set of chirped mirrors.
While in typical HHG experiments it is usual to pre-focus laser beam before the jet in order to optimize the phase-matching, in our experiment the beam was focused on the Argon jet in order to minimize the spot-size  of the driving laser on the jet.
Argon gas was injected into a vacuum chamber via a Parker's general pulsed valve with  a straight nozzle of 500 $\mu m$ and  with back pressure of 4.5 bar.

For our experimental conditions, the Keldysh parameter, $\gamma=0.63$, lies just outside the limiting value of the ADK formula, as discussed above, but is well within the accuracy range of the PPT formula.

\begin{figure}[thbp] 
    \begin{center}
	\includegraphics[width=\linewidth]{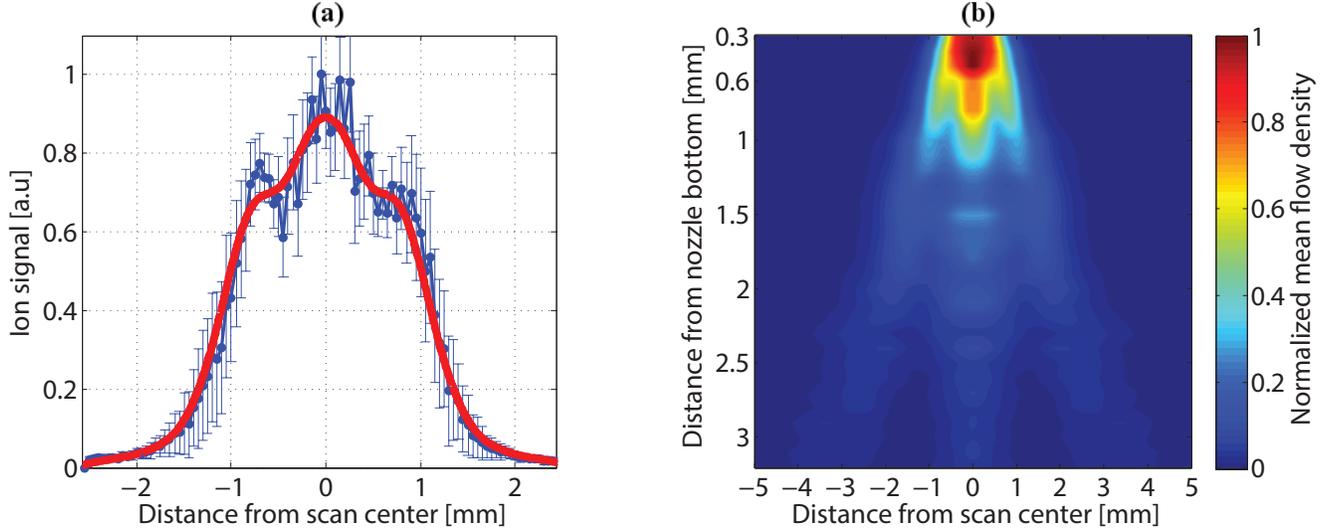}
    \captionsetup{justification=justified}

\caption{ Experimental results:
(a) Typical measured scan along Y axis (shown for X=0.9mm below nozzle).
(b) Reconstructed normalized mean flow density ($\rho_{MF}(r)$) shown across (XY) plane
as defined in Fig.  \ref{fig:system}.}
\label{Fig3_experiment}
\end{center}
\end{figure} %

The ion signal was measured, as a function of the jet position  in the XY plane (as defined in Fig. \ref{fig:system}).
A fast photodiode (PD) was used to monitor the pulse-to-pulse laser energy fluctuations, so that the measured ion signal could be calibrated using the PPT  model.
Fig. \ref{Fig3_experiment}\textcolor{blue}{a} shows a typical measured signal along the Y axis for a fixed X position (in blue).
The gas density $\rho(r,t)$, injected into the vacuum chamber from the gas jet, can be described as comprised of 2 contributions $\rho(r,t)= \rho_{MF}(r)+\delta\rho(r,t)$, where $\rho_{MF}(r)$ is the mean flow term \cite{Frumker_aerooptics_AO_2004}  and $\delta\rho(r,t)$ is the turbulent, pulse-to-pulse varying term. We attribute small deviations from  the rotational symmetry, which is expected for a cylindrically symmetric jet, to the turbulent   $\delta\rho(r,t)$ contribution and to the measurement noise. The error bars (in blue) in Fig. \ref{Fig3_experiment}\textcolor{blue}{a} account for these contributions.

In our proof-of-principle experiment we focus on the measurement of the inherently cylindrically symmetric mean flow $\rho_{MF}(r)$ part. To get a good estimate of this contribution and resolve the well known problem of the inverse Abel transform sensitivity to measurement noise in tomographic reconstructions \cite{Smith_Abel_inversion_JQSRT_1988, Dribinski_Abel_recosntruction_RSI_2002}, we symmetrized the measured data by calculating the average around the mean (first moment) for each measured scan and applied a low pass filter (LPF) to the result in order to minimize the impact of the measurement noise and $\delta\rho(r,t)$ density fluctuations. The Gaussian LPF, $f(\nu)\propto \exp\{-\nu^2/(2\sigma_{\nu}^2)\}$,  was applied in the frequency domain. The value of $\sigma_{\nu}$ varied for the different measured projections of the slices along the jet (with a typical value of $\sigma_{\nu}\simeq1100\textrm{m}^{-1}$), in accordance with the amount of noise accompanied to each measured projection. Naturally, a projection with smaller gas density was accompanied by relatively larger measurement noise.
The solid red  line in Fig. \ref{Fig3_experiment}\textcolor{blue}{a} shows the result of this procedure for the specific scan.

To reconstruct the  $\rho_{MF}(r)$ we applied the inverse Abel transform, implemented by the filtered back-projection algorithm, to the measured scans.
 To feed the measured data into the filtered back-projection (FBP) reconstruction algorithm, the spacing between the 'projection angles'  was chosen to be $0.1$ degrees apart. Thus, a total of $1800$ 'projection angles' were used in the back-projection part of the FBP algorithm.  Since the jet has cylindrical symmetry, for each 'angle of projection' the same data of the actually single measured projection was used.
 The resulting spatial density distribution of the gas jet is shown in Fig. \ref{Fig3_experiment}\textcolor{blue}{b}. This density distribution carries the signature of the exact shape of the nozzle and the poppet, the gas back pressure, temperature and its hydrodynamic properties, and the opening dynamics of the poppet. Specifically, there is a pre-pulse in the region of x=1.5mm, y=0, which can be an indication of the pre-opening behaviour of the poppet.

The trade-off between resolution, noise and measurement statistics is a well known, long standing problem in the area of Abel inversion tomographic techniques in the presence of measurement noise \cite{Dribinski_Abel_recosntruction_RSI_2002, Garcia_2D_charged_particle_image_Inv_RSI_2004, Smith_Abel_inversion_JQSRT_1988}.
In our proof-of-principle experiment, the measurement noise was mitigated by applying the LPF that effectively limits the resolution of the reconstruction below its fundamental limit.
In the future work, it will be important to improve the signal-to-noise ratio of the data acquisition, as well as to investigate the problem of finding the most efficient reconstruction algorithm within the context of strong field ionization tomography in order to realize its full potential up to the fundamental physical resolution limits, as presented in the theoretical section of this paper.

In general, theoretical and numerical calculations of the gas space-time density evolution exiting the jet is a highly non-trivial task and is a subject of the active research in the area of computational fluid dynamics (CFD) \cite{Smith_cryo_high_pressure_gas_jet_RSI_1998, Coussirat_CFD_gas_jet_JFE_2005, Luria_supersonic_beams_JPCA_2011}.
Our method opens an effective route for experimental validation, and therefore, a means for improving and refining of these gas jet dynamics' theoretical/computational models.

To provide absolute scale in our measurements, we can either theoretically \cite{miller1988free} evaluate
the density at a specific point within the jet, or calibrate our measurement experimentally with
a known density test target. Once calibrated, our approach would provide absolute spatial density
distribution within the measured target. In our experimental conditions, we estimate maximum density to be $\simeq 5 \times 10^{19} \textrm{cm}^{-3}$ just at the nozzle's exit with estimated $\sim10^{13}$ ions generated in the focal volume along the beam propagation at this point.

\section*{Conclusions}

We introduced laser induced strong-field tunneling gas jet tomography that features significant
resolution improvements when compared with the hitherto used techniques and enables targeted localized measurement in complex environments. We analysed its unique properties and presented a proof of principles experiment.  Our method, which is fully compatible with
typical HHG and other strong-field experimental set-ups, will allow simple yet robust gas jet density mapping in all those measurements. Attosecond science has long been restricted to spatially averaged measurements that obscure single molecule response to the strong field. Spatial mapping of the generating medium's density along with measurements of spectrally resolved wavefronts and complete space-time reconstruction of attosecond pulses \cite{frumker2012order} will
pave the way to much greater experimental accuracy in attosecond science.

However, the quest for accurate gas jet density characterization goes far beyond attosecond science.
Accurate spatial density characterization of gas jets is important for inertial confinement fusion (ICF) \cite{Denavit__gas_targets_PhysPlasmas_1994}, x-ray sources \cite{Failor_fluorescence_gas_tomogr_RSI_2003}, laser particle acceleration (LPA) \cite{Modena_Electron_nature_1995},  cold chemistry \cite{thorpe2009tomography} and other areas of science and engineering.
Hitherto only linear tomography modalities \cite{malka2000charact,Failor_fluorescence_gas_tomogr_RSI_2003} were up to the task.
 Our approach can be easily transferred to all these experiments.


\section*{Acknowledgements}

We thank Yuri Lyubarsky for stimulating discussions.
EF acknowledges support by the Israel Science Foundation
(grant No. 1116/14) and European Commission Marie Curie
Career Integration Grant.
MD, MN, SS and CV acknowledge financial support from the European Research Council Starting Research Grant UDYNI (Grant No. 307964) and from the Italian Ministry of Research and Education (ELI Grant  "ESFRI Roadmap").

%
%
%
%
%

\end{document}